\documentclass[%
aip,
rsi,
twocolumn,
amsmath,amssymb,
groupedaddress,
 reprint,%
]{revtex4-1}

\usepackage[utf8]{inputenc}
\usepackage[T1]{fontenc}

\usepackage{amsfonts,bm}
\usepackage{graphicx,epstopdf}  
\usepackage{color}
\usepackage{extarrows}
\usepackage{enumitem}
\usepackage{empheq}
 \usepackage{array}

\allowdisplaybreaks[1]

\newcommand{\cf}{cf.\,}

\renewcommand{\d}{{\rm d}}

\newcommand{\Sch}{Schr\"{o}dinger\ }

\newcommand{\w}{\omega}

\newcommand{\B}{\mbox{\tiny B}}

\newcommand{\tS}{\mbox{\tiny S}}

\newcommand{\T}{\mbox{\tiny T}}

\newcommand{\SB}{\mbox{\tiny SB}}

\newcommand{\dg}{\dagger}
\newcommand{\la}{\langle}
\newcommand{\ra}{\rangle}
\newcommand{\lla}{\langle\!\langle}
\newcommand{\rra}{\rangle\!\rangle}
\newcommand{\La}{\big\la}
\newcommand{\Ra}{\big\ra}

\newcommand{\Sec}[1]{Sec.\,\ref{#1}}
\newcommand{\nl}{\nonumber \\}
\newcommand{\be}{\begin{equation}}
\newcommand{\ee}{\end{equation}}
\newcommand{\bsube}{\begin{subequations}}
\newcommand{\esube}{\end{subequations}}
\newcommand{\Eq}[1]{Eq.\,(\ref{#1})}
\newcommand{\Eqs}[1]{Eqs.\,(\ref{#1})}
\newcommand{\Fig}[1]{Fig.\,\ref{#1}}

\usepackage{tikz}
\usetikzlibrary{positioning, shapes.geometric}
\newcommand{\RN}[1]{%
  \textup{\rm \uppercase\expandafter{\romannumeral#1}}%
}

\newcommand{\greater}{\mbox{\tiny $>$}}
\newcommand{\lesser}{\mbox{\tiny $<$}}
\newcommand{\lgter}{\mbox{\tiny $\lessgtr$}}
\newcommand{\gler}{\mbox{\tiny $\gtrless$}}

\begin{document}

\title{
Quantum mechanics of open systems: Dissipaton theories
}
\author{Yao Wang}
\author{YiJing Yan}
\email{yanyj@ustc.edu.cn}
\affiliation{University of Science and Technology of China, Hefei, Anhui 230026, China}

\date{\today}

 \begin{abstract}
%
%
This Perspective
presents a comprehensive account of the dissipaton theories developed in our group since 2014, including
the physical picture of dissipatons and the phase--space dissipaton algebra.  
The  dissipaton-- equation--of--motion--space (DEOM--space) formulations  cover the \Sch picture,  the Heisenberg picture, and further the imaginary--time DEOM.
Recently developed are also the dissipaton theories for studying equilibrium and nonequilibrium thermodynamic mixing processes. 
 The Jarzynski equality and Crooks relation are  accurately reproduced numerically.
It is anticipated that dissipaton theories  would remain essential towards a maturation of quantum mechanics of open systems.

\end{abstract}
\maketitle

\section{Introduction}
\label{Sec:Intro}
 Open quantum systems are ubiquitous in various fields of
science,\cite{Wei21,Kle09,Bre02,Yan05187}
covering quantum optics,\cite{Scu97, Lou73,Haa7398,Hak70,Sar74}
nuclear magnetic resonance,\cite{Rei82,Sli90,Van051037}
condensed matter and material physics,\cite{Bor85,Hol59325,Hol59343,Kli97,Ram98}
quark-gluon plasma,\cite{Aka15056002,Bla181,Miu20034011,Yao212130010}
nonlinear spectroscopy, \cite{Muk95,She84,Muk81509,Yan885160,Yan91179,Che964565,Tan939496,Tan943049}
chemical and biological physics. \cite{Nit06,Lee071462,Eng07782,Dor132746,Cre13253601,Kun22015101}
In all these studies, the total system--plus--bath composite Hamiltonian
assumes the form of $H_{\T}=H_{\tS}+h_{\B}+H_{\SB}$,
which together with temperature and/or chemical potentials
constitute a \emph{thermodynamic} system.
Irreversibility takes place, in terms of not only
relaxation, dephasing and quantum transport events, but also
those fundamental processes
subject to the Laws of Thermodynamics.

 In literature, quantum dissipation theories (QDTs), such as quantum master equations,\cite{Red651,Lin76393,Cal83587,Kam92,Xu029196}
focus mainly on the reduced system density operator,
$\rho_{\tS}(t)\equiv{\rm tr}_{\B}\rho_{\T}(t)$.
Exact QDTs include the Feynman--Vernon influence
functional path integral formalism \cite{Fey63118}
and its time--derivative equivalence, hierarchical equations of motion (HEOM), with either bosonic
\cite{Tan20020901,Tan89101,Tan906676,Yan04216,Tan06082001,Xu05041103,Xu07031107,Din12224103}
or fermionic bath environment influence.\cite{Jin08234703,Li12266403,Ye16608}
However, as mentioned earlier,
the reduced system ($H_{\tS}$) is just the primarily interested part
of the thermodynamic system that is characterized by
not only the total composite $H_{\T}$, but also temperatures and other
thermodynamic parameters.
The relevant information encoded in the reduced system dynamics
 alone would be insufficient to deal with
experimental measurements on open systems.
The entangled system--and--environment dynamics are also crucially important.
\cite{Imr02,Hau08}

In this paper, we present a systematic framework of quantum mechanics of open systems, in terms of dissipaton theories.
These are universal theories developed on the basis of both quantum mechanics and statistical thermodynamics principles.
In \Sec{thsec2}, we present the statistical quasi-particle picture of dissipatons, followed by the related algebraic constructions.\cite{Yan14054105,Yan16110306}
Adopted for the bath Hamiltonian ($h_{\B}$) is
the Gaussian environments ansatz  that is rooted 
at the well--known central limit theorem.
In other words, the bath effectively consists
of infinite number of noninteracting (quasi) particles, either bosonic or fermionic.
Consider further
the system--bath coupling $(H_{\SB})$, a superposition of $\{\hat Q_{u}^{\tS}\hat F_{u}^{\B}\}$, in which the  dissipative system  modes $\{\hat Q^{\tS}_u\}$ are arbitrary and the hybrid bath  modes $\{\hat F^{\B}_u\}$ assume linear.
This is the scenario of  Gauss--Wick's environments.  The bath influences are then completely characterized by the hybridization bath correlation functions.
Dissipatons can now be deduced for satisfying 
 the  generalized diffusion equation as required by the theory.
 Dynamical variables are
 dissipaton density operators (DDOs), whose time--evolutions are governed by the dissipaton equation of motion (DEOM).
 The reduced system density operator ($\rho_{\tS}$) is just a member of DDOs.
Another fundamental ingredient of dissipaton algebra is the generalized Wick's theorem. This enables dissipaton  theories for not only the reduced system but also the hybrid bath modes.
%

Section \ref{thsec3} comprises
 a complete description of DEOM--space quantum mechanics of open systems,
constructed in parallel to those traditional Liouville--space formulations.
 These include  the real--time dynamics in \Sch versus Heisenberg pictures, the imaginary--time dynamics, and  the DEOM evaluations on such as expectation values and correlation functions.

Section \ref{thsec4} is concerned with the thermodynamic mixing via the dissipaton implementations.
These include  the equilibrium $\lambda$-DEOM for the Helmholtz free--energy change \cite{Gon20154111} and the nonequilibrium $\lambda(t)$-DEOM for the work distributions.\cite{Gon22054109}
 The Jarzynski equality and Crooks relation are accurately
reproduced with numerical DEOM evaluations. \cite{Gon22054109}

Section \ref{thsec5} is concerned with the
dissipaton thermofield (DTF) theory. This covers the thermofield  dissipaton Langevin equation and  
 the nonequilibrium system--bath
entanglement theorem. Established are the relations between the  local system correlation functions and those involving the nonlocal hybrid bath modes.
In \Sec{thsecsum}, we discuss the future prospect of the dissipaton theories towards the quantum mechanics of open systems. 
%
%
%
Finally,  we conclude this paper.

\section{Onsets of dissipatons}\label{thsec2}

\subsection{Prelude}
 Let us start with the total system--plus--bath composite Hamiltonian
in the generic form of
\be\label{HSB_boson}
 H_{\T}=H_{\tS}+h_{\B}+H_{\tS\B}
 =H_{\tS} +\sum_{\alpha}h_{\alpha}
  +\sum_{\alpha u}\hat Q_{u}\hat F_{\alpha u}.
\ee
Both the system Hamiltonian $H_{\tS}$
 and the dissipative system modes $\{\hat Q_u\}$ are arbitrary,
including the time--dependence
via the classical external fields, which
act on the system and/or the neighboring environment.
For brevity, we set throughout this paper $\hbar=1$ and
$\beta_{\alpha}=1/(k_{B}T_{\alpha})$,
the inverse temperature of the $\alpha$--reservoir.

The hybrid reservoir bath modes $\{\hat F_{\alpha u}\}$
assume to be linear.
This together with noninteracting and
independent reservoir bath, $h_{\B}=\sum_{\alpha}h_{\alpha}$,
constitute the so--called Gaussian--Wick's
coupling environments.\cite{Wei21,Kle09}
 Their influences are \emph{fully} characterized by
the hybridization bath correlation functions,
satisfying the fluctuation--dissipation theorem, \cite{Wei21,Kle09,Bre02,Yan05187,Zhe121129}
\be\label{Fcorr_boson}
 \la\hat F^{\B}_{\alpha u}(t)\hat F^{\B}_{\alpha v}(0)\ra_{\B}
=\frac{1}{\pi}\!\int_{-\infty}^{\infty}\!\d\w\,
 \frac{e^{-i\w t}J_{\alpha uv}(\w)}{1-e^{-\beta_{\alpha}\omega}}.
\ee
Here, $\hat F^{\B}_{\alpha u}(t)\equiv
e^{ih_{\B}t}\hat F_{\alpha u}e^{-ih_{\B}t}
=e^{ih_{\alpha}t}\hat F_{\alpha u}e^{-ih_{\alpha}t}
$ and $\la(\,\cdot\,)\ra_{\B}\equiv{\rm tr}_{\B}[(\,\cdot\,)
\rho^{0}_{\B}]$, with $\rho^0_{\B}=\otimes_{\alpha} [e^{-\beta_{\alpha}h_{\alpha}}/{\rm tr}_{\B}
(e^{-\beta_{\alpha}h_{\alpha}})]$.
The involving hybridization bath spectral densities are
\be\label{Jw_def_boson}
 J_{\alpha uv}(\w) =\frac{1}{2i}\!\int_{-\infty}^{\infty}\!\d t\,e^{i\w t}
  \phi_{\alpha uv}(t)
\ee
with
\be\label{phit}
 \phi_{\alpha uv}(t)
\equiv i\la[\hat F^{\B}_{\alpha u}(t),\hat F^{\B}_{\alpha v}(0)]\ra_{\B}.
\ee

 To proceed, we decompose \Eq{Fcorr_boson}  into ($t\geq 0$)
\be\label{FDT_boson}
c_{\alpha uv}(t)
\equiv \la\hat F^{\B}_{\alpha u}(t)\hat F^{\B}_{\alpha v}(0)\ra_{\B}
\simeq \sum_{k=1}^K\eta_{\alpha uv k}e^{-\gamma_{\alpha k} t}.
\ee
This can be readily achieved with some sum--over--pole schemes
 \cite{Hu10101106,Hu11244106,Din11164107,Din12224103,Zhe121129}
or the time--domain Prony fitting decomposition
scheme.\cite{Che22221102}
For simplicity of formulations,
we assume all $\gamma_{\alpha uvk}=\gamma_{\alpha k}$ (valid at least when the poles of $J_{\alpha uv}(\w)$ contain no explicit system--mode dependence).
Let further ${\bar k}\in \{k=1,\cdots,K\}$
via $\gamma_{\alpha \bar k}\equiv \gamma^{\ast}_{\alpha k}$,
since the exponents $\{\gamma_{\alpha k}\}$
are either real or complex conjugate paired.
We can then report the required time--reversal relation,
$\la \hat F_{\alpha v}^{\B}(0)\hat F_{\alpha u}^{\B}(t)\ra_{\B}
=\la\hat F^{\B}_{\alpha u}(t)\hat F^{\B}_{\alpha v}(0)\ra^{\ast}_{\B}$,
in terms of
\begin{align}\label{FBt_corr_rev}
 \la \hat F_{\alpha v}^{\B}(0)\hat F_{\alpha u}^{\B}(t)\ra_{\B}
  =\!\sum_{k=1}^{K}\eta_{\alpha uv \bar k}^{\ast} e^{-\gamma_{\alpha k} t}.
\end{align}
Apparently, $\bar k=k$ if $\gamma_{\alpha k}$ is real,
since $\gamma_{\alpha \bar k}\equiv\gamma_{\alpha k}^{\ast}$.

\subsection{Dissipaton decomposition}

  The dissipaton theory provides a statistical quasi-particle (dissipaton)
picture to account for the environment,
starting with the dissipaton decomposition,
\be\label{hatFB_in_f}
 \hat F_{\alpha u}=\sum_{k=1}^{K}  \hat f_{\alpha u k},
\ee
where $\{\hat f_{\alpha u k}\}$ are known as dissipaton operator.
To reproduce the required \Eqs{FDT_boson} and (\ref{FBt_corr_rev}),
we set
\be\label{ff_corr_boson}
\begin{split}
 \la\hat f_{\alpha u k}^{\B}(t)\hat f_{\alpha' v k'}^{\B}(0)\ra_{\B}
=\delta_{\alpha \alpha'}\delta_{kk'}
  \eta^{\greater}_{\alpha u v k}e^{-\gamma_{\alpha k}t},
\\
 \la\hat f_{\alpha' v k'}^{\B}(0)\hat f_{\alpha u k}^{\B}(t)\ra_{\B}=\delta_{\alpha \alpha'}\delta_{kk'}
  \eta^{\lesser}_{\alpha uv k}e^{-\gamma_{\alpha k}t},
\end{split}
\ee
where
\be\label{eta_forback_def}
\begin{split}
&\eta_{\alpha uv k}^{\greater}\equiv
 \la \hat f_{\alpha  u k}\hat f_{\alpha v k}\ra^{\greater}_{\B}
 \equiv \la \hat f_{\alpha  u k}(0+)\hat f_{\alpha  v k}\ra_{\B}
=\eta_{\alpha uv k},
\\
&\eta_{\alpha uv k}^{\lesser}
\equiv \la \hat f_{\alpha  v k}\hat f_{\alpha  u k}\ra^{\lesser}_{\B}
 \equiv \la \hat f_{\alpha  v k}\hat f_{\alpha  u k}(0+)\ra_{\B}
=\eta^{\ast}_{\alpha uv \bar k}.
  \end{split}
\ee
Note that
$\la\hat f_{\alpha' v \bar k}^{\B}(0)
\hat f_{\alpha u \bar k}^{\B}(t)\ra_{\B}
=\la\hat f_{\alpha u k}^{\B}(t)\hat f_{\alpha' v k}^{\B}(0)\ra_{\B}^{\ast}$.

\begin{figure}[t]
\includegraphics[width=\columnwidth]{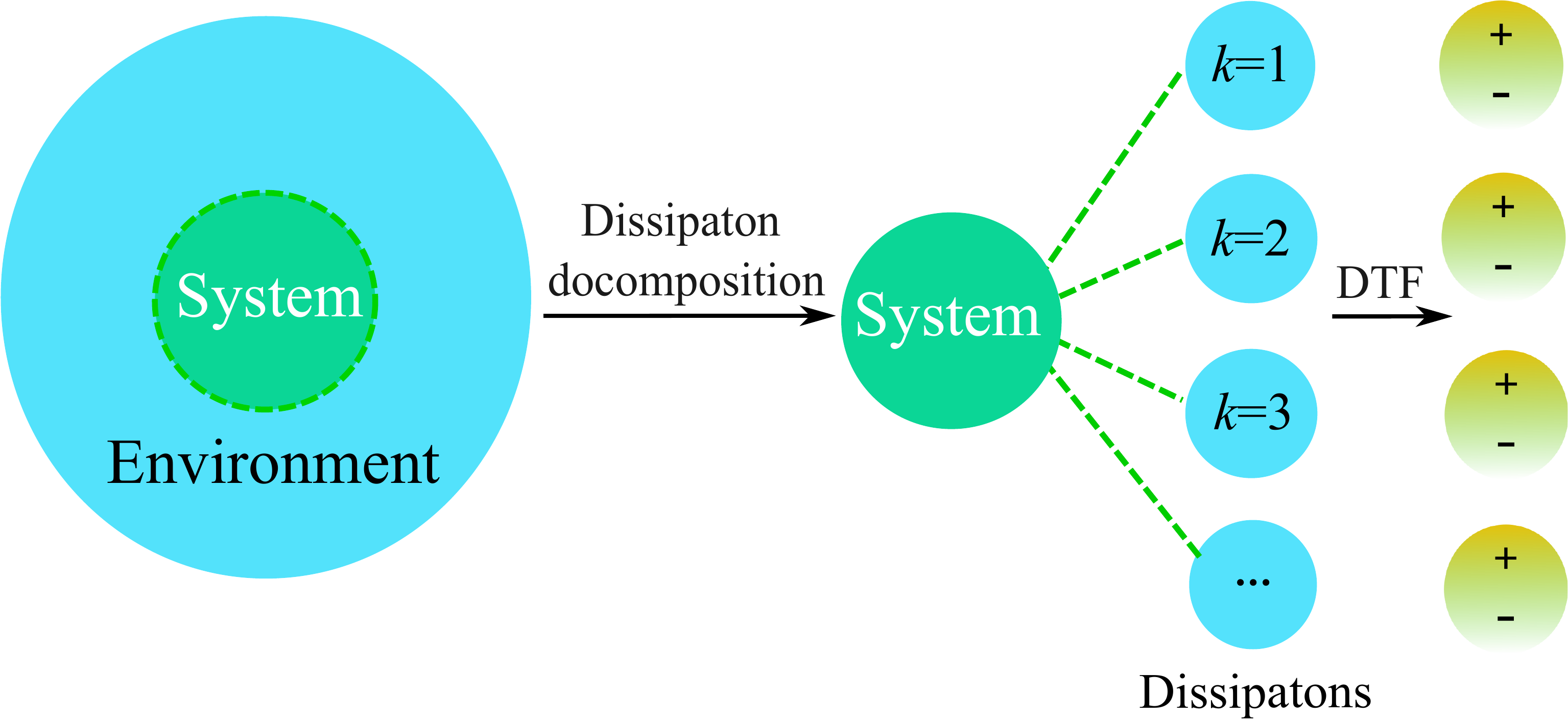}
\caption{Dissipaton decomposition [\cf\Eq{hatFB_in_f}] followed by the DTF decomposition. The latter will be  explained in \Sec{thsec5} [\cf\Eqs{tf1} and (\ref{van})].
}\label{fig1}
\end{figure}

The dissipaton decomposition,
\Eqs{hatFB_in_f}--(\ref{eta_forback_def}),
represented by the first mapping arrow in \Fig{fig1},
is concerned with individual $\hat F_{\alpha u}$
that assumes by far to be a linear operator
in the bare bath $h_{\alpha}$--subspace.
The resulting $\{\hat f_{\alpha u k}^{\B}(t)\equiv e^{ih_{\B}t}
\hat f_{\alpha u k}e^{-ih_{\B}t}\}$
are \emph{linear} and \emph{statistically independent}
diffusive environmental modes [\cf\Eq{ff_corr_boson}],
with the diffusion constant $\gamma_{\alpha k}$ that
can be complex.
	In other words,  \Eq{hatFB_in_f} essentially represents a mapping from the $\alpha$-reservoir bath to $K$ independent auxiliary baths, which intrinsically conserves the correlation functions in \Eqs{FDT_boson} and (\ref{FBt_corr_rev}).

  To proceed, we introduce
the \emph{irreducible} dissipaton product notation,
\be\label{circ_intro}
\big(\hat f_{k}\hat f_{j}\big)^{\circ}=\big(\hat f_{j}\hat f_{k}\big)^{\circ}.
\ee
This is true for bosonic dissipatons.
As \Eqs{hatFB_in_f}--(\ref{eta_forback_def})
reproduce the bosonic fluctuation--dissipation theorem,
\Eq{Fcorr_boson}, the dissipatons
$\{\hat f_{\alpha u k}\}$ in \Eq{hatFB_in_f}
are bosonic.
In this paper the dissipaton theories
are illustrated with the bosonic scenario.

\subsection{Disspaton density operators as dynamical variables}
\label{thsec2B}

 Dynamical variables in dissipaton theories
are the so-called dissipaton density operators (DDOs):\cite{Yan14054105,Zha15024112,Jin15234108}
\be \label{DDO}
  \rho^{(n)}_{\bf n}(t)
\equiv {\rm tr}_{\B}\Big[
  \Big(\prod_{\alpha u k} \hat f_{\alpha u k}^{n_{\alpha u k}}\Big)^{\circ}\rho_{\T}(t)
 \Big].
\ee
This describes a configuration that is irreducible
and labeled by an ordered collection of indexes, ${\bf n}\equiv\{n_{\alpha u k}\}$,
with $n_{\alpha u k}=0,1,2,\cdots$
being the participation number of individual
bosonic $\hat f_{\alpha u k}$-dissipaton.
The total number of dissipaton
excitations in $\rho^{(n)}_{\bf n}(t)$
is given by
\be\label{eq2.12}
  n=\sum_{\alpha u k} n_{\alpha u k}.
\ee
The reduced system density operator
is $\rho_{\tS}(t) = \rho^{(0)}_{\bm{0}}(t)$,
just a special member of DDOs.

 Let $\rho^{(n\pm 1)}_{{\bf n}^{\pm}_{\alpha u k}}$
be the associated $(n\pm 1)$-dissipatons configuration, with
${\bf n}^{\pm}_{\alpha u k}$ differing from ${\bf n}$ only
at the specified $\hat f_{\alpha u k}$-disspaton participation number,
$n_{\alpha u k}$, by $\pm 1$.
For presenting the related dissipaton algebra,
adopt hereafter the following notations:
\be\label{DDO_action}
\begin{split}
  \rho^{(n)}_{\bf n}(t;\hat A^{\times})\equiv
  {\rm tr}_{\B}\Big[\Big(\prod_{\alpha u k} \hat f_{\alpha u k}^{n_{\alpha u k}}\Big)^{\circ}
  \hat A^{\times}\rho_{\T}(t)\Big],
\\
 \rho^{(n)}_{\bf n}(t;\hat A^{\lgter})\equiv
  {\rm tr}_{\B}\Big[\Big(\prod_{\alpha u k} \hat f_{\alpha u k}^{n_{\alpha u k}}\Big)^{\circ}
  \hat A^{\lgter}\rho_{\T}(t)\Big],
\end{split}
\ee
where $\hat A^{\times}\equiv \hat A^{\greater}-\hat A^{\lesser}$,
\be\label{Algrter}
 \hat A^{\greater}\rho_{\T}(t)\equiv \hat A\rho_{\T}(t)
\ \ \text{and}\ \
 \hat A^{\lesser}\rho_{\T}(t)\equiv \rho_{\T}(t)\hat A .
\ee
The above notations will appear in the generalized diffusion
equation and the generalized Wick's theorem.
These are two fundamental ingredients of dissipaton algebra as follows; see also \Sec{thsec3}.

\subsection{Generalized diffusion equation}\label{thesec}
 Equation (\ref{ff_corr_boson}) highlights two important features of dissipatons:
(\emph{i}) Dissipatons with different ``color-$\gamma_{\alpha k}$''
are statistically independent with respective to the $(\alpha k)$-index;
(\emph{ii})
Each individual dissipaton goes by a \emph{single-exponential} correlation function,
with a \emph{same exponent} for both the forward and the backward paths.
These features are closely related to the dissipaton algebra
used in the DEOM construction.
 In particular, the feature (\emph{ii}) above leads to \cite{Yan14054105}
\be\label{dot_f_boson}
 {\rm tr}_{\B}\Big[\Big(\frac{\partial}{\partial t}
\hat f_{\alpha u  k}\Big)_{\B}\rho_{\T}(t)\Big]
 =-\gamma_{\alpha  k}\,
    {\rm tr}_{\B}\big[\hat f_{\alpha  u k}\rho_{\T}(t)\big] .
\ee
This is the \emph{generalized diffusion equation} for dissipatons.
It together with
 $(\frac{\partial}{\partial t} \hat f_{\alpha u  k})_{\B}
=i[h_{\B}, \hat f_{\alpha u  k}]
\equiv i h_{\B}^{\times}\hat f_{\alpha u  k}$
gives rise to the $h_{\B}^{\times}$--action on DDOs the result of
\be\label{gendiff}
 \rho^{(n)}_{\bf n}(t;h_{\B}^{\times})
=-i\Big(\sum_{\alpha u k} n_{\alpha uk} \gamma_{\alpha k}\Big)
  \rho^{(n)}_{\bf n}(t).
\ee

 Denote for bookkeeping later,
\be\label{eq17}
 H_0\equiv H_{\tS}+h_{\B}
\ \,\text{and}\ \,
 {\cal L}^{(n)}_{\bf n} \equiv H^{\times}_{\tS}
   -i\sum_{\alpha uk} n_{\alpha uk} \gamma_{\alpha k}\,.
\ee
Together with \Eq{gendiff}, we obtain
\be\label{eq18}
 \rho^{(n)}_{\bf n}(t;H_{0}^{\times})
={\cal L}^{(n)}_{\bf n}\rho^{(n)}_{\bf n}(t).
\ee

\subsection{Generalized Wick's theorems}

 Another important ingredient of dissipaton algebra
is the \emph{generalized Wick's theorem} (GWT). Consider first
\begin{align}\label{Wick_greater_boson}
&\quad\,\rho^{(n)}_{\bf n}(t;\hat f_{\alpha' vk'}^{\gler})
\equiv
 {\rm tr}_{\B}\Big[\Big(\prod_{\alpha  u k}
  \hat f^{n_{\alpha  u k}}_{\alpha  u k}\Big)^{\circ}
  \hat f^{\greater}_{\alpha'  v k'}\rho_{\T}(t)\Big]
\nl&
 =\sum_{\alpha  u k} n_{\alpha  u k}
  \la\hat f_{\alpha  u k}\hat f_{\alpha'  v k'}\ra^{\greater}_{\B}
  \rho^{(n-1)}_{{\bf n}^{-}_{\alpha  u k}}(t)+\rho^{(n+1)}_{{\bf n}^{+}_{\alpha'v k'}}(t).
\end{align}
Here, $\la \hat f_{\alpha  u k}\hat f_{\alpha'v k'}\ra^{\greater}_{\B}
 \equiv \la \hat f_{\alpha  u k}(0+)\hat f_{\alpha'  v k'}\ra_{\B}$
via \Eq{ff_corr_boson}, with the nonzero value being only
$\eta_{\alpha uv k}^{\greater}
\equiv\la \hat f_{\alpha  u k}\hat f_{\alpha v k}\ra^{\greater}_{\B}$.
The $\hat f^{\lesser}_{\alpha'  v k'}$--action counterpart
to \Eq{Wick_greater_boson} is similar,
but goes with
$\la \hat f_{\alpha'  v k'}\hat f_{\alpha  u k}\ra^{\lesser}_{\B}
\equiv \la \hat f_{\alpha'  v k'}\hat f_{\alpha  u k}(0+)\ra_{\B}$,
whose nonzero value is
$\eta_{\alpha uv k}^{\lesser}
\equiv \la \hat f_{\alpha  v k}\hat f_{\alpha  u k}\ra^{\lesser}_{\B}$
only.
We can then recast \Eq{Wick_greater_boson}
with the unified expression of
\be\label{w_x}
 \rho^{(n)}_{\bf n}(t;\hat f_{\alpha uk}^{\gler})
=\rho^{(n+1)}_{{\bf n}^{+}_{\alpha uk}}(t)
 +\!\sum_{v} n_{\alpha vk} \eta^{\gler}_{\alpha vuk}
  \rho^{(n-1)}_{{\bf n}^{-}_{\alpha vk}}(t).
\ee
This is the GWT for bosonic dissipatons,
which as seen later determines $\rho^{(n)}_{\bf n}(t;H_{\SB}^{\times})$,
the last two terms in the DEOM (\ref{DEOM}).
Involved are the pre--exponential coefficients
in \Eq{FDT_boson} or \Eq{FBt_corr_rev},
since $\eta_{\alpha uv k}^{\greater}=\eta_{\alpha uv k}$
and $\eta_{\alpha uv k}^{\lesser}=\eta^{\ast}_{\alpha uv{\bar{k}}}$,
as specified in \Eq{eta_forback_def}.

 It is worth re-emphasizing that the GWT,
\Eq{w_x}, goes by the irreducibility nature
of $\rho^{(n)}_{\bf n}(t)$: In \Eq{DDO},
the product of dissipaton operators inside $(\cdots)^\circ$
is irreducible, satisfying
the bosonic permutation relation of \Eq{circ_intro}.
In comparison, we may recall some properties about the
``normal order'' in textbooks,
which arranges creation operators
before annihilation operators.
Denote this also with $(\,\cdot\,)^{\circ}$,
such that $(\hat a^{\dg} \hat a )^{\circ}=(\hat a\hat a^{\dg})^{\circ}
=\hat a^{\dg} \hat a$.
The ground state $|0\ra$ satisfies $\hat a|0\ra=0$.
Set $\hat f \equiv \sqrt{\eta}\,(\hat a+\hat a^{\dg})$,
with $\eta$ being an arbitrary real parameter.
It is easy to obtain
$(\hat f^n)^{\circ}\hat f=(\hat f^{n+1})^{\circ}+n\eta(\hat f^{n-1})^{\circ}
=\hat f(\hat f^n)^{\circ}$.
Equation (\ref{w_x}) is just
the dissipaton generalization of this result.

 To complete phase--space dissipaton algebra,\cite{Wan20041102}
consider further
\be\label{momentum1}
\hat \Phi_{\alpha u}\equiv \dot {\hat F}_{\alpha u}
= i[h_{\alpha}, {\hat F}_{\alpha u}],
\ee
which satisfies
\be\label{momentum2}
 [\hat F_{\alpha u},\hat \Phi_{\alpha' v}]
=i\delta_{\alpha \alpha'}\Theta_{\alpha u v},
\ee
with
\be\label{diss_frequency}
 \Theta_{\alpha uv}\equiv \frac{1}{\pi}\!
  \int_{-\infty}^{\infty}\!{\rm d}\w\,\w J_{\alpha uv}(\w).
\ee
The dissipaton momentum decomposition is [\cf\Eq{hatFB_in_f}]
\be\label{pinphi}
  {\hat\Phi}_{\alpha u}=\sum_{k=1}^{K}\hat\varphi_{\alpha u k}.
\ee
The resulting GWT for dissipaton momentums
reads\cite{Wan20041102}
\begin{align}\label{w_p}
 \rho^{(n)}_{\bf n}(t;\hat\varphi_{\alpha uk}^{\gler})
&=\gamma_{\alpha k}\sum_{v} n_{\alpha  vk} \eta_{\alpha vuk}^{\gler}
  \rho^{(n-1)}_{{\bf n}^{-}_{\alpha vk}}(t)
\nl& \quad \,
 -\gamma_{\alpha k}\rho^{(n+1)}_{{\bf n}^{+}_{\alpha uk}}(t).
\end{align}
Introduce now
\be\label{psipm_def}
 \hat\theta^{\pm}_{\alpha u k}
\equiv \frac{1}{2}
 \Big(\hat f_{\alpha u k}\mp
   \frac{\hat \varphi_{\alpha u k}}{\gamma_{\alpha k}}
 \Big).
\ee
We obtain \Eqs{w_x} and (\ref{w_p}) the alternative
expressions,
\be\label{Wick_Xu}
\begin{split}
 \rho^{(n)}_{\bf n}(t;\hat\theta^{+;\gler}_{\alpha u k})
&=\rho^{(n+1)}_{{\bf n}^{+}_{\alpha u k}}(t),
\\
 \rho^{(n)}_{\bf n}(t;\hat\theta^{-;\gler}_{\alpha u k})
&=\sum_{v} n_{\alpha  v k} \eta^{\gler}_{\alpha vu k}
  \rho^{(n-1)}_{{\bf n}^{-}_{\alpha  v k}}(t).
\end{split}
\ee

\section{DEOM--space quantum mechanics}
\label{thsec3}

\subsection{Time-evolutions of DDOs}
\label{thsec3A}
The construction of DEOM starts with 
\be \label{Lioueq}
\dot\rho_{\T}(t)=-i[H_{0}+H_{\tS\B},\rho_{\T}(t)],
\ee
where $H_{0}\equiv H_{\tS}+h_{\B}$ and
[cf.\,\Eq{hatFB_in_f}]
\be
 H_{\tS\B}=\sum_{\alpha u}\hat Q_{u}\hat F_{\alpha u}
=\sum_{\alpha uk}\hat Q_{u} \hat f_{\alpha u k}.
\ee
Applying \Eq{Lioueq} to \Eq{DDO} and further \Eqs{eq18}--(\ref{w_x}), we obtain
\begin{align}\label{DEOM}
 \dot\rho^{(n)}_{\bf n}
&=-i{\cal L}^{(n)}_{\bf n}\rho^{(n)}_{\bf n}
 -i\sum_{\alpha uk}
  \hat Q_{u}^{\times}\rho^{(n+1)}_{{\bf n}_{\alpha u k}^+}
\nl&\quad\,
  -i\!\sum_{\alpha uvk}n_{\alpha uk}
  \big(\eta_{\alpha u v k}\hat Q_{v}^{\greater}
      -\eta_{\alpha u v \bar k}^{\ast}\hat Q^{\lesser}_{v}
  \big)\rho^{(n-1)}_{{\bf n}_{\alpha uk}^-}.
\end{align}
In parallel to $\dot\rho_{\T}(t)=-i{\cal L}_{\T}\rho_{\T}(t)$ [\cf\Eq{Lioueq}],
we can recast the set of linear equations  (\ref{DEOM}) as 
\be\label{sdeom}
\dot{\bm\rho}(t)=-i\bm{\mathcal{L}}{\bm\rho}(t),
\ \, \text{with}\ \,
 {\bm \rho}(t)\equiv \{\rho_{\bf n}^{(n)}(t)\}.
\ee
The dynamical generator $\bm{\mathcal{L}}$,
defined via \Eq{DEOM}, can be time--dependent in general.
In line with $\rho_{\T}\rightarrow {\bm\rho}$
and ${\cal L}_{\T}\rightarrow\bm{\mathcal{L}}$,
the DEOM--space formulations are
those of the Liouville--space mappings, as follows.

Evidently, DEOM (\ref{DEOM}) is identical to the well--established HEOM formalism.\cite{Tan20020901,Tan89101,Tan906676,Yan04216,Tan06082001,Xu05041103,Xu07031107,Din12224103}
All numerical methods developed for HEOM, illustrated in \Sec{3C}, are applicable in DEOM evaluations.  Now the observables cover not only the reduced system but also the hybrid bath properties [\cf\Eq{adeom}]; see also 
\Sec{thsec4} for  the dissipaton theory implementations of equilibrium and nonequilibrium thermodynamics.
It is worth emphasizing that the underlying dissipaton algebra can be readily extended to nonlinear coupling environments.\cite{Xu17395,Xu18114103,Che22Arxiv2206_14375}
This scenario is beyond the conventional HEOM approach that is rooted at  the  Feynman--Vernon influence functional path integral formalism.

\subsection{DEOM--space observables}
\label{thsec3B}

 Consider the expectation values,
\begin{align}\label{expectation}
\bar A(t)&\equiv {\rm Tr}[\hat A \rho_{\T}(t)]\equiv\lla\hat A|\rho_{\T}(t) \rra
\nl &
=\lla\hat {\bm A}|{\bm\rho}(t) \rra\equiv \sum_{\bf n}{\rm tr}_{\tS}\big[\hat A_{\bf n}^{(n)}\rho_{\bf n}^{(n)}(t)\big].
\end{align}
The second line denotes the DEOM--space evaluation,
which as inferred from the \Eqs{w_x} and (\ref{w_p}),
supports the following types of operators,
\be\label{adeom}
\hat A\in \{\hat A_{\tS}, \hat B_{\tS}\hat f_{\alpha u k}, \hat B_{\tS}\hat \varphi_{\alpha u k}\}.
\ee
Here, $\hat A_{\tS}$ and $\hat B_{\tS}$ are
arbitrary observables in the system subspace, including
$\hat B_{\tS}=\hat I_{\tS}$,
whereas
$\{\hat f_{\alpha u k}\}$ and
$\{\hat \varphi_ {\alpha u k}\}$ are related to hybrid bath modes $\{\hat F_{\alpha u}\}$  and  $\{\hat \Phi_{\alpha u}\equiv \dot {\hat F}_{\alpha u}\}$ via \Eqs{hatFB_in_f} and (\ref{pinphi}), respectively.
To complete \Eq{expectation},
we map $\hat A$ into the DEOM--space operators,
\be\label{mapping}
\hat A \rightarrow \hat {\bm A}\equiv\{\hat A_{\bf n}^{(n)}; n=0,1,2,\cdots\}.
\ee
The dissipaton algebra established earlier results in
\bsube\label{DEOM_map}
\begin{align}\label{smap}
\hat A_{\tS}&\rightarrow \hat {\bm A}=\{{\hat A^{(0)}=\hat A_{\tS}; \ \ \hat A_{\bf n}^{(n>0)}=0}\},
\\
\label{sxmap}
\hat B_{\tS}\hat f_{\alpha u k}&\rightarrow \hat {\bm A}=\{{\hat A^{(1)}_{\alpha u k}=\hat B_{\tS}; \ \ \text{others}=0}\},
\\  \label{spmap}
\hat B_{\tS}\hat \varphi_{\alpha u k}&\rightarrow \hat {\bm A}=\{{\hat A^{(1)}_{\alpha u k}=-\gamma_{\alpha k}\hat B_{\tS}; \ \ \text{others}=0}\}.
\end{align}
\esube
We can then evaluate the expectation values for these types of operators
via the last identity of \Eq{expectation}.

 Turn to the steady--state correlation functions,
which in general can be recast as [cf.\,\Eq{expectation}]
\be\label{corr_hilbert}
\la \hat A(t) \hat B(0)\ra
=\lla \hat A|\rho_{\T}(t;\hat B^{\greater})\rra
=\lla \hat {\bm A}|{\bm\rho}(t;\hat B^{\greater})\rra,
\ee
with $\rho_{\T}(t;\hat B^{\greater})
= e^{-i{\cal L}_{\T}t}\rho_{\T}(0;\hat B^{\greater})
\equiv e^{-i{\cal L}_{\T}t}(\hat B\rho_{\T}^{\rm st})$
and
\be
 \rho_{\T}(t;\hat B^{\greater})
\rightarrow {\bm\rho}(t;\hat B^{\greater})
\equiv\{\rho_{\bf n}^{(n)}(t;\hat B^{\greater})\}.
\ee
Both $\hat A$ and $\hat B$ belong to the types of \Eq{adeom}.
Moreover, in relation to $\rho_{\T}(0;\hat B^{\greater})
\equiv \hat B\rho_{\T}^{\rm st}$,
the initial values of DDOs in
evaluating \Eq{corr_hilbert} are given by
\be\label{DDO_initial0}
 \rho^{(n)}_{\bf n}(0;\hat B^{\greater})\equiv
  {\rm tr}_{\B}\Big[
   \Big(\prod_{\alpha u k} \hat f_{\alpha u k}^{n_{\alpha u k}}\Big)^{\circ}
  \hat B^{\greater}\rho^{\rm st}_{\T}\Big].
\ee
For the first type operator of \Eq{mapping}, we have
\bsube\label{DDO_initial}
\be
  \rho^{(n)}_{\bf n}(0;\hat B_{\tS}^{\greater})
=\hat B_{\tS}\rho^{(n);{\rm st}}_{\bf n}.
\ee
The other two types are related to [\cf\Eq{Wick_Xu}]
\be\label{Wick_Xu_initial}
\begin{split}
\rho^{(n)}_{\bf n}(0;\hat B^{\greater}_{\tS}\hat\theta^{+;\greater}_{\alpha u k})
&=\hat B_{\tS}\rho^{(n+1)}_{{\bf n}^{+}_{\alpha u k}}(t),
\\
 \rho^{(n)}_{\bf n}(0;\hat B^{\greater}_{\tS}\hat\theta^{-;\greater}_{\alpha u k})
&=\sum_{v} n_{\alpha  v k} \eta_{\alpha vu k}
  \hat B_{\tS}\rho^{(n-1)}_{{\bf n}^{-}_{\alpha  v k}}(t),
\end{split}
\ee
\esube
since $\hat f_{\alpha u k}
= \hat\theta^{-}_{\alpha u k}+\hat\theta^{+}_{\alpha uk}$
and $\hat\varphi_{\alpha u k}
=\gamma_{\alpha k}(\hat\theta^{-}_{\alpha u k}-\hat\theta^{+}_{\alpha uk})$;
\Eq{psipm_def}.

 The DEOM evaluations of correlation functions are as follows:
(\emph{i}) Compute the steady--state DDOs, $\{\rho_{\bf n}^{(n);{\rm st}}\}$;
(\emph{ii}) Determine the initial values
$\{\rho^{(n)}_{\bf n}(0;\hat B^{\greater})\}$ via
applicable \Eq{DDO_initial} in study;
(\emph{iii}) Propagate DDOs with \Eq{DEOM} to obtain the required $\{\rho^{(n)}_{\bf n}(t;\hat B^{\greater})\}$;
(\emph{iv}) Evaluate $\la \hat A(t) \hat B(0)\ra$ as the expectation value problem by using \Eqs{expectation}--(\ref{DEOM_map}). Demonstrated examples of these evaluations include such
as Fano interferences,\cite{Zha15024112,Zha18780}  Herzberg--Teller vibronic spectroscopy \cite{Zha16204109,Che21244105}
and transport current noise spectrum.\cite{Wan20041102,Jin20235144,Mao21014104}

\subsection{DEOM in the Heisenberg picture}

 The Heisenberg picture of DEOM is concerned with
\[
 \hat A(t)=\hat Ae^{-i{\cal L}_{\T}t}
\rightarrow  \hat{\bm A}(t)=\hat{\bm A}e^{-i\bm{\mathcal{L}}t}
\equiv \{\hat A^{(n)}_{\bf n}(t)\},
\]
satisfying
\be 
\dot{\hat{\bm A}}(t)=-i\hat{\bm A}(t)\bm{\mathcal{L}}
\ee
and
\be\label{Aave_Hei}
  \la\hat A(t)\ra = \lla\hat{\bm A}(t) \big{|} {\bm\rho}(0)\rra
 = \lla\hat{\bm A}(0) \big{|} {\bm\rho}(t)\rra
\ee
with
$\hat{\bm A}(t=0) \equiv \hat{\bm A}$.
From $\la\!\la \dot{\hat{\bm A}}|{\bm\rho}\ra\!\ra
 = \la\!\la \hat{\bm A}|\dot{\bm\rho}\ra\!\ra$, 
we obtain
\begin{align}\label{DEOM_bose_Hei}
  \dot{\hat A}^{(n)}_{\bf n} &=
  -i\hat A^{(n)}_{\bf n}{\cal L}^{(n)}_{\bf n}
  -i\sum_{\alpha u k} {\hat A}^{(n-1)}_{{\bf n}^{-}_{\alpha u k}}
   \hat Q_u^{\times}
  -i\!\sum_{\alpha u v k}(n_{\alpha u k }+1)
\nl&\qquad\times
  {\hat A}^{(n+1)}_{{\bf n}^{+}_{\alpha u k}}
   (\eta_{\alpha uvk}\hat Q^{\greater}_{v}
  -\eta^{\ast}_{\alpha uv\bar k}\hat Q^{\lesser}_{v}).
\end{align}
This is the Heisenberg picture counterpart to \Eq{DEOM},
where $\hat O\hat B^{\greater}=\hat O\hat B$
and $\hat O \hat B^{\lesser}= \hat B \hat O$,
in line with \Eq{Algrter}.

The main usage of \Eq{DEOM_bose_Hei}
is concerned with efficient evaluations of
nonlinear correlation functions,
such that
 $\la\hat A(t_2)\hat B(t_1)\hat C(0)\ra
=\lla \hat{\bm A}(t_2)|\hat{\bm B}|{\bm\rho}(t_1;\hat C^{\greater})\rra$,
via the mixed Heisenberg--\Sch DEOM dynamics. 
The formulation here is closely related to the doorway--window picture of pump--probe spectroscopy.\cite{Yan906485}
As also known, the pump can be an  optimal control field, whereas the probe provides a means of feedback.

\subsection{DEOM toolkits and related considerations}\label{3C} 
As mentioned earlier,  the DEOM (\ref{DEOM}) itself is identical to the well--established HEOM. \cite{Tan06082001,Xu05041103,Xu07031107,Din12224103}
Various methods developed there can be directly applied; see the recent Perspective by Y.\ Tanimura.\cite{Tan20020901}
New developments include the follows: (\emph{i}) The adiabatic terminator for hierarchy level truncation, which alleviates the numerical long--time instability problems;\cite{Zha21905} (\emph{ii}) The time--domain Prony fitting decomposition scheme for accurate and minimum dissipaton basis set, applicable to arbitrary hybridization bath spectral densities;\cite{Che22221102} (\emph{iii}) The implementation of matrix product state;\cite{Shi18174102, Yan21194104,Ke22194102,Tuc66279,Kol09455,Hac09706,Gra1353} (\emph{iv})  The transformed Brownian oscillator basis; \cite{Yan20204109,Li22064107} (\emph{v}) The construction of rate kernels via DEOM by utilizing the Nakajima--Zwanzig
projection techniques \cite{Zha163241,Su224554}
and so on.

%

In the following, we focus on the steady--state solver and related imaginary--time DEOM formalism.

\subsubsection{Efficient steady--state solver}

Steady states play crucial roles in many equilibrium and non-equilibrium open system studies, including aforementioned correlation function problems.
The standard choices for solving high-dimension linear equations
are the Krylov subspace methods.\cite{Saa03}
Nevertheless, solving the steady states DDOs, $\dot{\bm \rho}^{\rm st}=0$ or $\bm \rho(t\rightarrow \infty)$, via HEOM/DEOM (\ref{DEOM}) is often a challenging task, since the vast number of dynamical quantities are involved.
The proposed self--consistent iteration (SCI) approach would be the choice.\cite{Zha17044105}
%
%
To be concrete, we set $\dot\rho^{(n)}_{\bf n}=0$ and obtain
\begin{align*}
  0
&=-i{\cal L}^{(n)}_{\bf n}\rho^{(n);{\rm st}}_{\bf n}
 -i\sum_{\alpha uk}
  \hat Q_{u}^{\times}\rho^{(n+1);{\rm st}}_{{\bf n}_{\alpha u k}^+}
\nl&\quad\,
  -i\!\sum_{\alpha uvk}n_{\alpha uk}
  \big(\eta_{\alpha u v k}\hat Q_{v}^{\greater}
      -\eta_{\alpha u v \bar k}^{\ast}\hat Q^{\lesser}_{v}
  \big)\rho^{(n-1);{\rm st}}_{{\bf n}_{\alpha uk}^-}.
\end{align*}
Then recast it into the SCI equation,\cite{Zha17044105}
\begin{align}\label{SCI}
 \rho^{(n);{\rm st}}_{\bf n}
&=
( i{\cal L}^{(n)}_{\bf n}+\epsilon)^{-1}\bigg[\epsilon\rho^{(n);{\rm st}}_{\bf n} -i\sum_{\alpha uk}
  \hat Q_{u}^{\times}\rho^{(n+1);{\rm st}}_{{\bf n}_{\alpha u k}^+}
\nl&\quad\,
  -i\sum_{\alpha uvk}n_{\alpha uk}
  \big(\eta_{\alpha u v k}\hat Q_{v}^{\greater}
      -\eta_{\alpha u v \bar k}^{\ast}\hat Q^{\lesser}_{v}
  \big)\rho^{(n-1);{\rm st}}_{{\bf n}_{\alpha uk}^-}\bigg] 
\end{align}
where $\epsilon>0$ is 
an arbitrary parameter.
The SCI evaluation is  subject to the constraint $\text{tr}_{\tS}\rho^{(0)}_{\bf 0}=1$.
The iteration will converge as long as
the diagonal part of $( i{\cal L}^{(n)}_{\bf n}+\epsilon)$ dominates.
Increasing $\epsilon$ will increase the numerical stability,
but decrease the convergence speed.
For a good balance between accuracy and efficiency,
it is appropriate to have $\epsilon$ the value about
the spectrum span of the system Hamiltonian.

As known, the SCI equation (\ref{SCI}) accommodates the hierarchical structure and the efficient on-the-fly filtering algorithm.\cite{Shi09084105}
The numerical practices 
also show the remarkable 
advantages of SCI scheme over the Krylov subspace methods.\cite{Zha17044105}

\subsubsection{Imaginary-time DEOM ($i$-DEOM)}\label{sec:ideom}
Alternatively, the equilibrium state can be related to the imaginary--time DEOM, \cite{Gon20154111} which aims at hybridization partition function,
\be \label{uu2}
 Z_{\rm hyb}\equiv Z_{\T}/Z_{0}
\equiv {\rm Tr}\varrho_{\T}(\beta),
\ee
with $Z_{\T}\equiv{\rm Tr}\,e^{-\beta H_{\T}}$ and $Z_{0}(T)\equiv{\rm Tr}\,e^{-\beta H_{0}}$.
Only single bath is involved so that the $\alpha$-index  is dropped.
Evidently, $Z_{\rm hyb}$ is related to the hybridization free--energy that can also be evaluated via the $\lambda$--thermodynamic integral; \cf\Sec{thsec4A}.
The imaginary--time dynamics is concerned with
\be\label{varrhoT_tau}
 \varrho_{\T}(\tau)=
 e^{-\tau H_{\T}}e^{-(\beta-\tau)H_{0}}\big/Z_0,
\ee
which satisfies [cf.\,\Eq{Algrter}]
\be\label{iLiou_eq}
 \frac{\d}{\d\tau}{\varrho}_{\T}(\tau)
=-\big(H^{\times}_{0} +H^{\greater}_{\tS\B}\big)\varrho_{\T}(\tau).
\ee
The $i$-DEOM--space mappings then go  by
\be
 \varrho_{\T}(\tau)\rightarrow
{\bm\varrho}(\tau) \equiv \{\varrho^{(n)}_{\bf n}(\tau)\},
\ee
with the $i$-DDOs satisfying
\be\label{iDEOMa}
 \frac{\d}{\d\tau}\varrho^{(n)}_{\bf n}(\tau)
=-\varrho^{(n)}_{\bf n}(\tau;H^{\times}_{0})
-\sum_{uk}\hat Q_{u}\varrho^{(n)}_{\bf n}(\tau; \hat{f}_{uk}^{\greater}).
\ee
In parallel to \Eqs{eq18} and (\ref{w_x}), we obtain\cite{Gon20154111}
\be\label{iDEOMb}
\begin{split}
 \varrho^{(n)}_{\bf n}(\tau;H^{\times}_{0})
&={\cal L}^{(n)}_{\bf n}\varrho^{(n)}_{\bf n}(\tau),
\\
 \varrho^{(n)}_{\bf n}(\tau; \hat{f}_{uk}^{\greater})
&=\varrho^{(n+1)}_{{\bf n}_{uk}^{+}}(\tau)
  +\sum_{v} n_{vk}\eta_{vuk}
   \varrho^{(n-1)}_{{\bf n}^{-}_{vk}}(\tau).
\end{split}
\ee
In line with $\varrho_{\T}(0)=e^{-\beta H_{0}}/Z_0$ of \Eq{varrhoT_tau},
the initial values of $i$-DDOs are
\be\label{iDDOs0}
\varrho^{(0)}_{\bf 0}(0)=e^{-\beta H_{\tS}}/Z_{\tS}
\ \ \text{and} \ \
 \varrho_{\bf n}^{(n>0)}(0)=0,
\ee
where $Z_{\tS}={\rm tr}_{\tS}e^{-\beta H_{\tS}}$.
We then propagate the $i$--DEOM  until $\tau=\beta$ and evaluate
\be 
 Z_{\rm hyb}={\rm tr}_{\tS}\varrho^{(0)}_{\bf 0}(\beta).
\ee
%
It can be further verified that \cite{Gon20154111}
\be \label{equ}
\frac{\varrho^{(n)}_{\bf n}(\beta)}{{\rm tr}_{\tS}[\varrho^{(0)}_{\bf 0}(\beta)]}=\rho_{\bf n}^{(n);{\rm eq}}.
\ee
The right--hand--side are the equilibrium DDOs, which can be obtained via steady--state solvers, such as the SCI [\cf\Eq{SCI}].

\section{Dissipaton implementations of thermodynamic mixing}\label{thsec4}

The system--bath entanglement plays crucially important roles in not only dynamics but also the
 thermodynamic properties. \cite{Kir35300,Gon20154111,Gon20214115}
The latter has just been illustrated with the $i$-DEOM
formalism.
In the following, 
we will introduce an alternative approach
and further extend  this method to the nonequilibrium scenarios.

\subsection{Equilibrium $\lambda$-DEOM formalism}
\label{thsec4A}
%

%

The equilibrium $\lambda$-DEOM focuses  on the free--energy change before and after isotherm system--bath mixing,
\be\label{Ahyb_0}
 A_{\rm hyb}(T)\equiv A_{\T}(T)- A_0(T).
\ee
Evidently, $A_{\rm hyb}(T)=-\beta^{-1}\ln Z_{\rm hyb}(T)$ [\cf\Eq{uu2}].
%
%
According to the thermodynamic integral formalism, the hybridization free--energy can be expressed as\cite{Kir35300,Shu71413,Gon20214115,Zon08041103,Zon08041104,Gon20154111}
\be \label{integration}
A_{\rm hyb}(T)=\int_{0}^{1}\frac{{\rm d}\lambda}{\lambda}\la H_{\SB} \ra_{\lambda}
\ee
where $\lambda$ is the mixing parameter and
\be\label{HSB_lambda_ave}
 \la  H_{\tS\B}\ra_{\lambda} \equiv
 {\rm Tr}[(\lambda H_{\tS\B})\rho^{\rm eq}_{\T}(T;\lambda)]
\ee
with $\rho^{\rm eq}_{\T}(T;\lambda) =e^{-\beta H_{\T}(\lambda)}/Z_{\lambda}(T)$ and  $Z_{\lambda}(T)\equiv{\rm Tr}\,e^{-\beta H_{\T}(\lambda)}$ [cf.\,\Eq{Hlam}].
The involving total composite Hamiltonian reads
\be \label{Hlam}
 H_{\T}(\lambda) = H_{0}  + \lambda H_{\tS\B},
\ee
with $H_{0}\equiv H_{\tS}+h_{\B}$ [\cf\Eq{eq17}].
%

 Equation (\ref{HSB_lambda_ave}) implies
that $\la H_{\tS\B}\ra_{\lambda}$ is just the $\lambda$--augmented
equivalence to the original $\la H_{\tS\B}\ra$ where $\lambda =1$.
As seen from \Sec{thsec3}, DEOM supports accurate evaluations of $\la H_{\tS\B}\ra_{\lambda}$ for the Gauss--Wick's bath.
In particular, for $H_{\SB}=\sum_u\hat Q_{u}\hat F_{u}$,
by using \Eqs{expectation} with (\ref{sxmap}), we obtain
\be\label{HSB_DEOM}
 \la H_{\tS\B}\ra_{\lambda}
=\lambda\sum_{uk}{\rm tr}_{\tS}\big[\hat Q_{u}
  \rho^{\rm eq}_{uk}(T;\lambda)\big].
\ee
Here, $\rho^{\rm eq}_{uk}(T;\lambda)\equiv
\rho^{(1);{\rm eq}}_{{\bf 0}^{+}_{uk}}(T;\lambda)$ is
a $\lambda$--augmented DDO at thermal equilibrium,
with the generic form of \Eq{DDO} but
\be \label{rhon_eq_lambda}
  \rho^{(n);{\rm eq}}_{\bf n}(T;\lambda)
\equiv {\rm tr}_{\B}\Big[
  \Big(\prod_{ u k} \hat f_{u k}^{n_{u k}}\Big)^{\circ}\rho^{\rm eq}_{\T}(T;\lambda)
 \Big].
\ee
The hybridization free--energy in \Eq{Ahyb_0} can be then obtained via the integration of $\lambda$ by using  \Eq{integration}. This is the equilibrium $\lambda$-DEOM formalism.
Practically,
 we evaluate ${\bm\rho}^{\rm eq}(T;\lambda)
\equiv \big\{\rho_{\bf {n}}^{(n);{\rm eq}}(T;\lambda)\big\}$
progressively, by noting that
$\rho^{(0);{\rm eq}}_{\bf 0}(T;0)=e^{-\beta H_{\tS}}/Z_{\tS}$
and $\rho_{\bf {n}}^{(n>0);{\rm eq}}(T;0)=0$.
Then use the known ${\bm\rho}^{{\rm eq}}(T;\lambda)$
as the initial values for calculating
${\bm\rho}^{{\rm eq}}(T;\lambda+\delta\lambda)$
via either the real--time ($t\rightarrow\infty$) propagation
or the steady--state solver.
We have also developed the free--energy spectrum theory for thermodynamics of open quantum impurity systems, which relates the thermodynamic spectral functions to the local impurity properties.\cite{Gon20214115} 


\subsection{Nonequilibrium $\lambda(t)$-DEOM formalism}\label{sec:neqtherm}

Turn to the isotherm mixing processes that are irreversible.
A time--dependent mixing function $\lambda(t)$,  subject to
$
 \lambda(t=0)=0$ and $\lambda(t=t_f)=1
$, 
would be used instead.
This represents  nonequilibrium scenarios in general, where the work distribution $p(w)$ is the key quantity in related studies.
There are the Jarzynski equality,
\cite{Jar11329}
\be \label{Jar}
 \La e^{-\beta w}\Ra\equiv\int_{-\infty}^{\infty}\!\!{\rm d} w \,e^{-\beta w} p(w)=e^{-\beta A_{\rm hyb}}
\ee
and the Crooks relation,
 \cite{Cro992721}
\be\label{CrooksEq}
p(w) = e^{\beta (w-A_{\rm hyb})} \bar{p}(-w).
\ee
The latter is about a pair of conjugate processes, with the forward  and backward processes being controlled by $\lambda(t)$ and $\bar\lambda(t)\equiv\lambda(t_f-t)$, respectively.\cite{Tal07F569}
%
%
Evidently,  the forward work distribution  $p(w)$ and the backward $\bar p(-w)$ cross at the point of reversible work, $w=A_{\rm hyb}$ that can be obtained via the equilibrium $\lambda$-DEOM formalism.

The nonequilibrium $\lambda(t)$--DEOM is a viable means to the accurate evaluation of 
$p(w)$.
The formulations start with  $H_{0}|n\ra=H_{\T}(\lambda=0)|n\ra=\varepsilon_n|n\ra$ and $H_{\T}(\lambda=1)|N\ra=E_N|N\ra$ before and after mixing.
The distribution of mixing work is given by\cite{Tal07050102}
\be\label{forwardfW}
 p(w) = \sum_{N,n}\delta(w - E_N + \varepsilon_n)
  P_{N,n}(t_f, 0)P_n(0).
\ee
Here,
$P_{n}(0)=e^{-\beta \varepsilon_n}/{Z_0}$
is the initial probability distribution,
and $P_{N,n}(t,0)=\big|\la N|\hat U_{\T}(t) |n \ra\big|^2$
is the transition probability with the propagator
$U_{\T}(t)$ being governed by  the Hamiltonian $H_{\T}(t)=H_{0}+\lambda(t)H_{\SB}$.
%
%
The non-zero $\dot \lambda(t)$ that describes the irreversibility engages in
\begin{align}\label{Vpm}
\hat V_{\pm}(t;\tau) 
=\exp_{\pm}\left[\frac{i\tau}{2}\!\int_{0}^{t}{\rm d}t'\, \dot{\lambda}(t')H_{\SB}\right],
\end{align}
which participates in the
work generating operator,\cite{Tal07050102, Sak21033001}

\be\label{Phitau1}
 \hat W_{\T}(t;\tau) =
 \hat U_{\T}(t)\hat V_{+}(t;\tau) \rho^{\rm eq}_{0}(T)\hat V_{-}(t;\tau)\hat U_{\T}^{\dg}(t).
\ee
It can be shown that\cite{Tal07050102,Sak21033001}
\be\label{varPhitau1}
p(w)=\frac{1}{2\pi}\!\int_{-\infty}^{\infty}\!\!{\rm d} \tau \,e^{-i w\tau } \big\{{\rm Tr}[\hat W_{\T}(t_f;\tau)]\big\}.
\ee
%
%

\begin{figure}
\centering
\begin{minipage}[t]{0.49\columnwidth}
\centering
\includegraphics[width=\columnwidth]{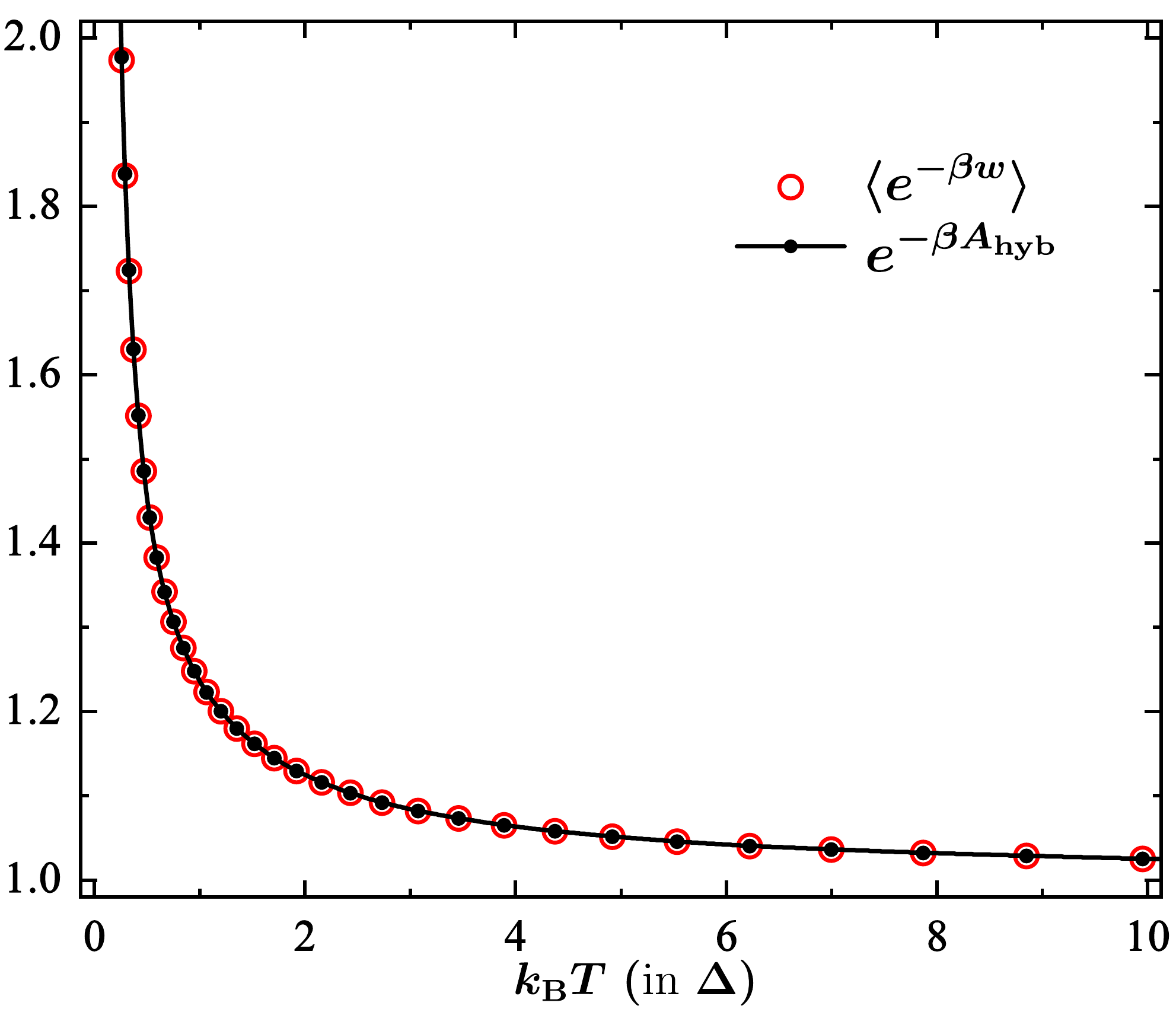}
\end{minipage}
\begin{minipage}[t]{0.49\columnwidth}
\centering
\includegraphics[width=0.995\columnwidth]{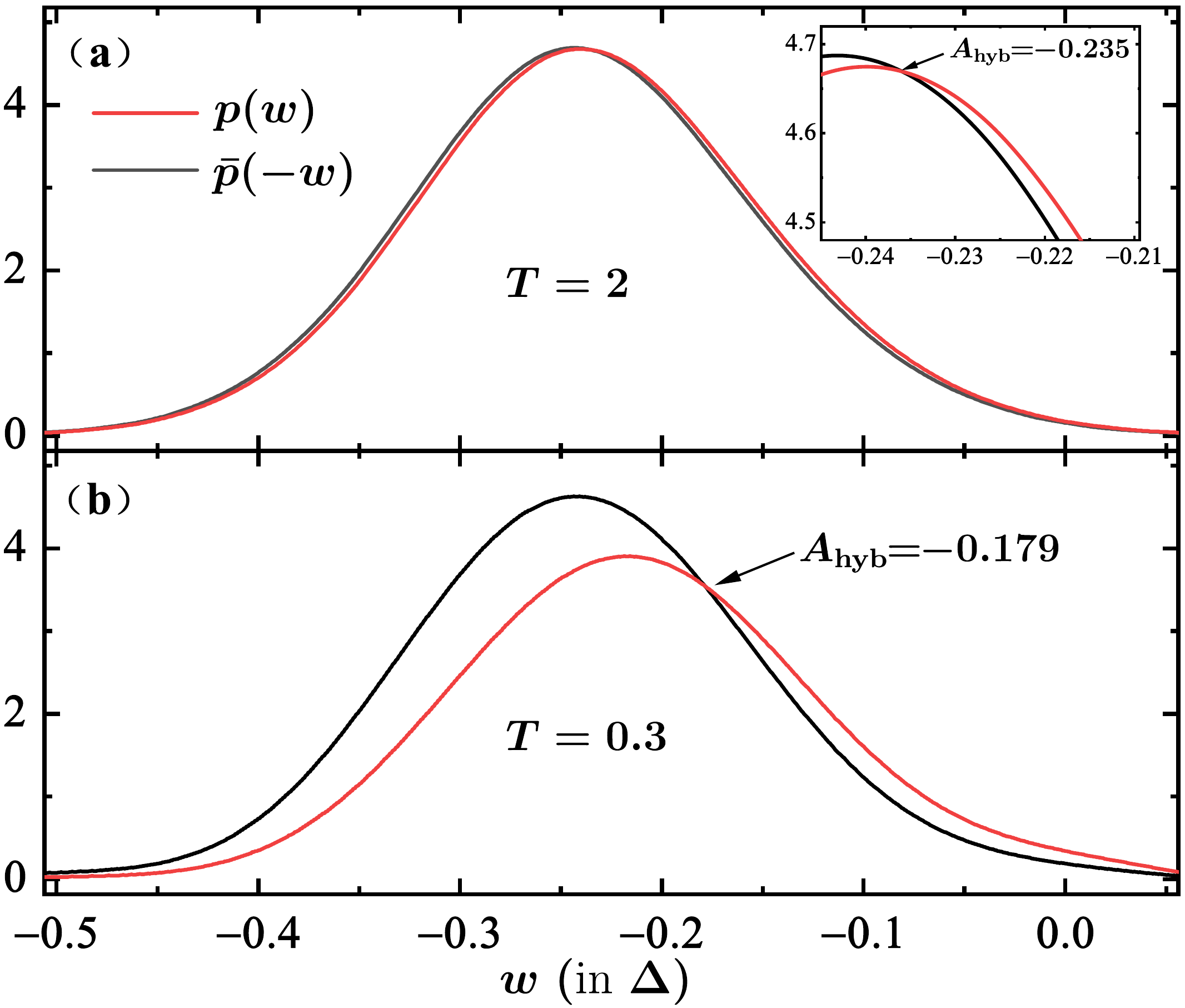}
\end{minipage}
\caption{Jarzynski equality (left panel) and Crooks relation (right panel), cited from Ref.\,\onlinecite{Gon22054109}. Numerical demonstrations are carried out with a spin-boson model,
in which the system Hamiltonian and dissipative mode are
$
  H_{\tS}={\varepsilon}\hat{\sigma}_z+\Delta\hat\sigma_x
\ \ \,\text{and}\ \ \,
\hat{Q}=\hat{\sigma}_{z},
$
respectively.
 Adopt for the bath spectral density the Drude model,
$
J(\w)=\eta\gamma\w/(\w^2+\gamma^2).
$
In all the simulations, we set $\varepsilon=0.5\Delta$,
$\gamma=4\Delta$ and $\eta=0.5\Delta$.
The forward time--dependent  mixing  function adopts
$
\lambda(t) = (1- e^{-\alpha t})/(1- e^{-\alpha t_f}),
$
with $\alpha=0.01\Delta$ and $t_f=50\Delta^{-1}$.
Simulations are carried out at different temperatures. In the left panel $T$ ranges from near $0$ to $10\Delta$, while it adopts 
(a) $T=2\Delta$ and (b) $T=0.3\Delta$ in the right panel.}\label{fig2}
\end{figure}

Turn to the equation of motion for the work generating operator $\hat W(t;\tau)$.
Its dynamics can be obtained as
\begin{align}\label{eom}
\frac{\partial \hat W_{\T}}{\partial t} 
&=-i[H_{0}^{\times}+\lambda_{-}(t) H_{\SB}^{\greater}
 -\lambda_{+}(t) H_{\SB}^{\lesser}]\hat W_{\T},
\end{align}
with
$
  \lambda_{\pm}(t)\equiv \lambda(t)\pm(\tau/2)\dot\lambda(t).
$
Initially, $\hat W_{\T}(0;\tau)=\rho^{\rm eq}_{0}(T)=e^{-\beta H_{0}}/Z_0$, as inferred from  \Eq{Phitau1}.
Similar to DDOs in \Eq{DDO},
we introduce the dissipatons--augmented
work generating operators (D-WGOs),
\be
 \hat W_{\T}(t;\tau)\rightarrow
\hat{{\bm W}}(t;\tau) \equiv \{\hat W^{(n)}_{\bf n}(t;\tau)\}.
\ee
The same procedure from \Eq{Lioueq} to \Eq{DEOM} now gives rise to \Eq{eom} the D-WGO correspondence,\cite{Gon22054109}
\begin{align}\label{DEOM_ss}
\frac{\partial {\hat W}^{(n)}_{\bf n}}{\partial t}&=-i{\cal L}^{(n)}_{\bf n}{\hat W}^{(n)}_{\bf n}
-i\sum_{uk} {\cal A}_{u}(t)\hat W_{{\bf n}_{uk}^+}^{(n+1)}
\nl & \quad
-i\sum_{uk} n_{uk}{\cal C}_{uk}(t) \hat W_{{\bf n}_{uk}^-}^{(n-1)},
\end{align}
where
\be
\begin{split}
&{\cal A}_u (t) \equiv
  \lambda_{-}(t) \hat Q^{\greater}_{u}
  -\lambda_{+}(t)  \hat Q^{\lesser}_{u},
 \\ &{\cal C}_{uk}(t) \equiv \sum_v \big [\lambda_{-}(t) \eta_{uvk} \hat Q^{\greater}_{v}
   -\lambda_{+}(t)\eta_{uv\bar k}^{\ast}
   \hat Q^{\lesser}_{v}\big].
   \end{split}
\ee
In relation to $\hat W_{\T}(0;\tau)=\rho^{\rm eq}_{0}(T)=e^{-\beta H_{0}}/Z_0$, the initial values to \Eq{DEOM_ss} are 
\be\label{iDDOs0}
\hat W^{(0)}_{\bf 0}(0;\tau)=e^{-\beta H_{\tS}}/Z_{\tS}
\ \ \text{and} \ \
 \hat W_{\bf n}^{(n>0)}(0;\tau)=0.
\ee
Finally, in line with \Eq{varPhitau1}, we evaluate
\be\label{pww}
p(w)=\frac{1}{2\pi}\int_{-\infty}^{\infty}\!\!{\rm d} w \,e^{-i w\tau } {\rm tr}_{\tS}[\hat W^{(0)}_{\bf 0}(t_f;\tau)].
\ee
Figure \ref{fig2}  reports the results in terms of 
the Jarzynski equality (\ref{Jar}) and the Crooks relation (\ref{CrooksEq}); see the figure caption for the details of model system.

\section{Dissipaton thermofield theory}\label{thsec5}

\subsection{Prelude}
Generally speaking, thermofield theory\cite{Ume95} is an important ingredient for quantum mechanics of open systems  and closely related to nonequilibrium Green's function formalisms.\cite{Sch61407,Kel651018,Wan14673}
The dissipaton thermofield theory (DTF)  \cite{Wan22044102}
to be presented in this section comprises 
in particular the
nonequilibrium system--bath entanglement theorem. This gives rise to relations between the  local system correlation functions and those involving the nonlocal hybrid bath modes.
The development exploits    further  decomposition of each dissipaton into the
 absorptive ($+$) and  emissive ($-$) components, as schematically  represented in the last column of \Fig{fig1}.

To proceed, we consider the $H_{\T}$--based Heisenberg picture
of the hybrid bath modes. It is easy to obtain \cite{Du20034102, Du212155}
\be\label{QLE_boson}
   \hat F_{\alpha u}(t)=\hat F^{\B}_{\alpha u}(t)
 -\sum_v \int^{t}_{0}\!\d\tau\,\phi_{\alpha uv}(t-\tau)\hat Q_v(\tau).
\ee
This is the  precursor to conventional quantum Langevin equation, with $\hat F^{\B}_{\alpha u}(t)$
being the random force and the related $\phi_{\alpha uv}(t)$ of \Eq{phit}.
%
%
Together with $[\hat F^{\B}_{\alpha u}(t), \hat Q_{v}(0)]=0$, we obtain the system--bath entanglement
theorem for response functions, which is a type of input--output relations in the total composite space.\cite{Du20034102,Du212155}

On the other hand, $\la\hat F^{\B}_{\alpha u}(t) \hat Q_{v}(0)\ra\neq 0$.
Equation (\ref{QLE_boson}) is insufficient to obtain the   input--output relations between nonequilibrium correlation functions, such as
\bsube\label{Cinput_boson_0}
\begin{align}
\label{Cinput_boson}
 {\bm C}_{\tS\tS}(t)=&\{C^{\tS\tS}_{uv}(t)\equiv \la \hat Q_{u}(t)\hat Q_v(0)\ra\},
 \\ \label{CalpS}
 {\bm C}_{\alpha\tS}(t)
= &\{C^{\alpha\tS}_{uv}(t)
 \equiv\la \hat F_{\alpha u}(t)\hat Q_{v}(0)\ra\}.
\end{align}
\esube
More specifically,  from \Eq{QLE_boson} we have
\be\label{CSalp_boson_final}
 {\bm C}_{\alpha \tS}(t)
={\bm X}_{\alpha\tS}(t) -\int_{0}^{t}\!{\rm d}\tau
\,{\bm \phi}_{\alpha}(t-\tau){\bm C}_{\tS\tS}(\tau),
\ee
with
\be
{\bm X}_{\alpha\tS}(t)=\{
X^{\alpha\tS}_{uv}(t)\equiv \la\hat F^{\B}_{\alpha u}(t)\hat Q_{v}(0)\ra\},
\ee
to be further resolved.
To that end, we  exploit the statistical quasi-particles picture, which is used in the DEOM theory \cite{Yan14054105}, and obtain
\be\label{termI1}
 {\bm X}_{\alpha\tS}(t)
 =\sum_k {\bm X}_{\alpha\tS k}(t),
\ee
with
\be \label{g2}
X_{uvk}^{\alpha \tS}(t)\!\equiv\! \la  \hat f^{\B}_{\alpha u k}(t)\hat Q_v(0)\ra\!=\!\la  \hat f_{\alpha u k}(0)\hat Q_v(0)\ra e^{-\gamma_{\alpha k}t}.\!
\ee
Here, $\{\gamma_{\alpha k}\}$ originates from the exponential decomposition of the interacting bath correlations as in \Eq{FDT_boson}.
Evidently, to establish the aforementioned correlation function type input--output relations, the key step is to formulate $X_{uvk}^{\alpha \tS}(0)$ in terms of ${\bm C}_{\tS\tS}(t)$ [cf.\,\Eq{Cinput_boson}] and ${\bm c}_{\alpha}(t)$ [cf.\,\Eqs{Fcorr_boson} and (\ref{FDT_boson})].
We address this issue within the scope of DTF theory to be elaborated as follows.

\subsection{Ansatzes}
The proposed DTF theory is based on
the dissipaton decomposition of the hybrid reservoir modes,
as schematically represented in \Fig{fig1}.
 There are three basic ingredients:

\noindent(\emph{i})
 \emph{Dissipaton decomposition ansatz:} The hybrid reservoir modes can be decomposed into dissipatons, as in \Eq{hatFB_in_f} with \Eq{ff_corr_boson}.

\noindent(\emph{ii})
 \emph{Thermofield dissipatons ansatz}: Each $\hat f_{\alpha u k}$ consists of
an absorptive ($+$) and an emissive ($-$) parts (\cf\Fig{fig1}),
\bsube\label{eq10}
\be \label{tf1}
\hat f_{\alpha uk} = \hat f_{\alpha uk}^{+}+\hat f_{\alpha uk}^{-},
\ee
defined via
\be\label{van}
 \hat f_{\alpha u k}^{-}\rho^0_{\B}
=\rho^0_{\B}\hat f_{\alpha u k}^{+}=0.
\ee
\esube
This results in
\be
\begin{split}
&c_{\alpha uv k}^{-}(t)
\equiv \la\hat f_{\alpha u k}^{-;\B}(t)\hat f_{\alpha v k}^{+;\B}(0)\ra_{\B}=\eta^{\greater}_{\alpha u v k}e^{-\gamma_{\alpha k}t},
\\
&c_{\alpha uv k}^{+}(t) \equiv \la\hat f_{\alpha v k}^{-;\B}(0)\hat f_{\alpha u k}^{+;\B}(t)\ra_{\B}=\eta^{\lesser}_{\alpha uv \bar k}e^{-\gamma_{\alpha k}t}.
\end{split}
\ee
As the thermofield excitation is concerned,
$\hat f_{\alpha u k}^{\pm}$ resembles
the creation/annihilation operator
onto the reference
$\rho^{0}_{\B}=\otimes_{\alpha} [e^{-\beta_{\alpha}h_{\alpha}}/{\rm tr}_{\B}
(e^{-\beta_{\alpha}h_{\alpha}})]$.\cite{Ume95}

\noindent(\emph{iii})
 \emph{Thermofield Langevin ansatz}:
Each thermofield dissipaton satisfies
\be\label{QLE2}
\hat f_{\alpha u k}^{\pm}(t)=\hat f_{\alpha u k}^{\pm;\B}(t)
 \pm i\sum_{v}\!\int^{t}_{0}\!\d\tau
  c^{\pm}_{\alpha uvk}(t-\tau)\hat Q_v (\tau).
\ee
In compared with \Eq{QLE_boson}, the resolved are not only the absorptive versus emissive contributions, but also the Langevin force that reads $\hat f_{\alpha uk}^{\pm;\B}(t)=\hat f_{\alpha uk}^{\pm;\B}(0) e^{-\gamma_{\alpha k} t}$.
This recovers the generalized diffusion equation of the DEOM theory. \cite{Yan14054105}

\subsection{System--bath entanglement theorem for correlation functions}
In the following, we elaborate above basic ingredients of the DTF theory, with a class of input--output relations between local and nonlocal nonequilibrium steady--state correlation functions.
Denote ${\bm C}_{\alpha\tS k}(t)=\{C_{\alpha \tS k}(t)\equiv \la \hat f_{\alpha u k}(t)\hat Q_{v}(0)\ra\}$ and ${\bm \phi}_{\alpha k}(t)=i[{\bm c}^{-}_{\alpha k}(t)-{\bm c}^{+}_{\alpha k}(t)]=\{\phi_{\alpha uvk}(t)= i[c_{\alpha uv k}^{-}(t)-c_{\alpha uv k}^{+}(t)]\}$.
Equations (\ref{eq10})
and (\ref{QLE2}) give rise to
\be\label{ft_O_corr}
{\bm C}_{\alpha\tS k}(t)={\bm X}_{\alpha \tS k}(t)
-\int_{0}^t\!{\rm d}\tau\,{\bm \phi}_{\alpha k}(t-\tau)
  {\bm C}_{\tS\tS}(\tau)
\ee
where $X_{uvk}^{\alpha \tS}(t)=X_{uvk}^{\alpha \tS}(0)e^{-\gamma_{\alpha k}t}$ [\cf \Eq{g2}] and
\be\label{fO_ave}
{\bm X}_{\alpha \tS k}(0)
  \!=i\!\int_{0}^{\infty}\!\!{\rm d}\tau
   \big[{\bm c}_{\alpha k}^{+}(\tau) {\bm C}_{\tS\tS}^{\dg}(\tau)
    - {\bm c}_{\alpha k}^{-}(\tau) {\bm C}_{\tS\tS}^{T}(\tau)\big].
\ee
Derivation details of \Eq{fO_ave} are shown in the next paragraph. Here, ${\bm M}^{T}$ denotes the matrix transpose of ${\bm M}$.
Together with \Eq{ft_O_corr}, we obtain further
\be\label{Xt_boson}
 {\bm X}_{\alpha\tS}(t)
=2\,{\rm Im} \! \int^{\infty}_0\!\!\d\tau\,
  {\bm c}_{\alpha}^{T}(t+\tau){\bm C}_{\tS\tS}^{T}(\tau).
\ee
This completes  \Eq{CSalp_boson_final},
the system--bath entanglement theorem for nonequilibrium steady--state correlation functions.

The derivations of the key expression (\ref{fO_ave}) are as follows.
(\emph{i}) Start with $\la \hat A(0)\ra
=\lim_{t\rightarrow\infty}{\rm Tr}\big[\hat A(t)\rho_{\T}^{\rm init}]$ for any operator $\hat A$.
This asymptotic identity holds
for any physically supported
initial total composite density
operator $\rho_{\T}^{\rm init}$.
In particular, we choose
$\rho_{\T}^{\rm init}=\rho_{\tS}^{\rm init}\otimes\rho^0_{\B}$, with $\rho^0_{\B}$ being the pure bath canonical ensemble density operator;
(\emph{ii}) Split $X_{uvk}^{\alpha \tS}(0)\equiv \la\hat f_{\alpha u k}(0)\hat Q_{v}(0)\ra=\la\hat f^{+}_{\alpha u k}(0)\hat Q_{v}(0)\ra+\la\hat Q_{v}(0)\hat f^{-}_{\alpha u k}(0)\ra$. This is true since the system and reservoir operators are  commutable  at any given local time;
(\emph{iii}) Obtain ${\rm Tr}[\hat f^{+}_{\alpha u k}(t)\hat Q_{v}(t)\rho_{\T}^{\rm init}]$
and ${\rm Tr}[\hat Q_{v}(t)\hat f^{-}_{\alpha u k}(t)\rho_{\T}^{\rm init}]$  from \Eq{QLE2},
with focus on their $t\rightarrow \infty$ expressions, where $\hat f^{\B;\pm}_{\alpha u k}(t)$ makes no contribution according to \Eq{van}.
The resulting
 $X_{uvk}^{\alpha \tS}(0)$ according to Step (\emph{ii})
is just \Eq{fO_ave}.

\subsection{Comments}
It is worth emphasizing that the DTF formalism, is rather general in relation to the absorptive and emissive processes. Its application to obtain \Eqs{ft_O_corr}--(\ref{Xt_boson}) is an example that can be numerically verified by DEOM evaluations. However, \Eqs{ft_O_corr}--(\ref{Xt_boson}) can not be obtained within the original DEOM framework.
That is to say, although both the DTF theory and DEOM method are numerically exact for Gaussian environments, DTF theory helps reveal more explicit relations.
Furthermore, the $t=0$ behaviour of \Eq{CSalp_boson_final} with \Eq{Xt_boson}
is closely related to
the nonequilibrium Green's function formalism of transport current. \cite{Sch61407,Kel651018,Mei922512,Hau08,Gru1624514} For example, consider
the heat transport from the $\alpha$--reservoir
to the local impurity system.  The heat current operator reads
\be\label{hatI_boson_def}
 \hat J_{\alpha}
\equiv
 -\frac{{\rm d}h_{\alpha}}{{\rm d}t}=-i[H_{\T},h_{\alpha}]
 =\sum_{u}\dot{\hat F}_{\alpha u}\hat Q_{u}.
\ee
This is the electron transport analogue. \cite{He18195437,Son17064308}
The heat current is then [cf.\,\Eq{CalpS}]
\be\label{heat_current}
 J_{\alpha}=\sum_{u}\la\dot{\hat F}_{\alpha u}\hat Q_{u}\ra
=\sum_{u}\dot{C}^{\alpha\tS}_{uu}(t=0).
\ee
Now apply \Eq{CSalp_boson_final}, with noticing
that its second term does not
contribute to $\dot{C}^{\alpha\tS}_{uu}(0)$.
We obtain \cite{Du212155}
\be\label{heat_current_final}
 J_{\alpha}=2\,{\rm Im}\!\int^{\infty}_0\!\!\d \tau\,
  {\rm tr}\big [\dot{\bm c}_{\alpha}(\tau){\bm C}_{\tS\tS}(\tau)\big].
\ee
This is the time--domain  Meir--Wingreen's formula. \cite{Mei922512} 
%
%

The DTF theory  would be better physically
supported than the conventional thermofield approach.
The latter involves the purification of bare bath canonical thermal states onto effective zero--temperature environments.\cite{Ume95}
On the other hand, the DTF approach exploits
the of discrete Brownian quasi-particle picture as implied in the dissipaton Langevin equation (\ref{QLE2}). This not only avoids introducing the auxiliary degrees of freedom, but also goes with the effectively resolved
random force, in line with 
the generalized diffusion equation (\ref{dot_f_boson}) of the DEOM
theory.
Moreover, the system--bath entanglement theorem for the properties of \Eq{Cinput_boson_0} can be readily extended to those such as $\la \hat F_{\alpha u}(t)\hat F_{\alpha' v}(0)\ra$ and also $\la \hat F_{\alpha u}(t)\hat A_{\tS}(0)\ra$ via $\la \hat Q_{u}(t)\hat A_{\tS}(0)\ra$ for an  arbitrary system operator $\hat A_{\tS}$.
%

Additionally, it is also worth noting that \Eq{QLE_boson}
immediately results in \cite{Gon20214115}
\be \label{FtoQ}
 \la\hat F_{\alpha u}\ra = -\sum_{v}\theta_{\alpha uv} \la\hat Q_v\ra,
 \ee
with
$
 \theta_{\alpha uv} \equiv \int^{\infty}_0\!\d t\, \phi_{\alpha uv}(t)
$ [{\rm cf}.\,\Eq{phit}]. 
This result can also be obtained 
via the DEOM formulations.

\section{Concluding and prospective remarks}\label{thsecsum}

\begin{figure}[t]
\includegraphics[width=0.9\columnwidth]{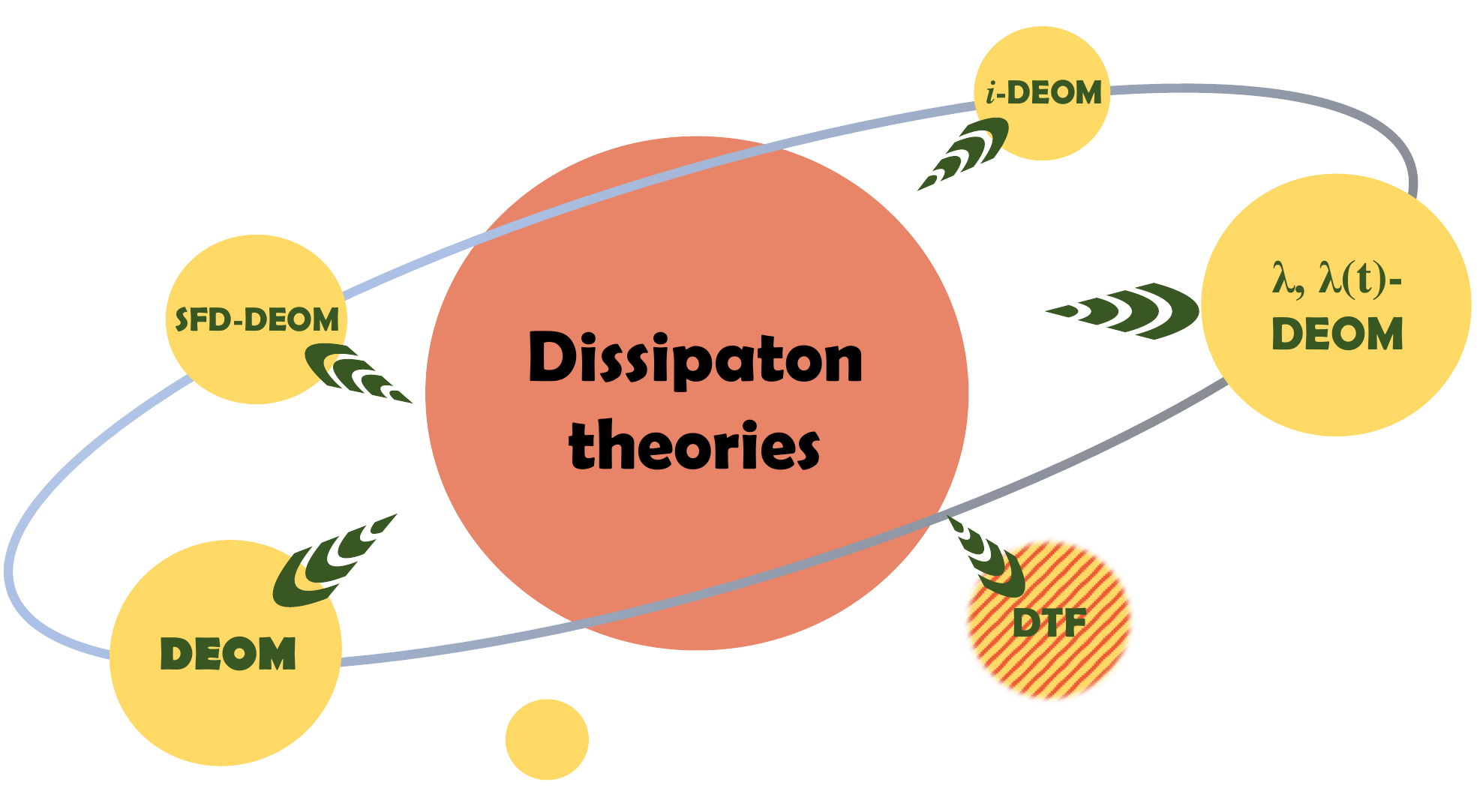}
\caption{The family of dissipaton theories, including now  real--time DEOM, imaginary--time DEOM ($i$-DEOM), equilibrium $\lambda$-DEOM, nonequilibrium $\lambda(t)$-DEOM, stochastic--field--dressed DEOM (SFD-DEOM) and the dissipaton thermofield (DTF) theory. Future family members are anticipated.
}\label{fig3}
\end{figure}

To conclude, 
dissipaton theories are essential building blocks of quantum mechanics of open systems,
comprising rich ingredients. These include the Brownian quasi-particles picture and the phase--space dissipaton algebra,  together with the dynamical variables (\Sec{thsec2}). The resulting real--time DEOM (\ref{DEOM}) unambiguously support various formulations such as expectation values, correlation functions, the Heisenberg picture,  the imaginary--time DEOM and so on (\Sec{thsec3}).
The $\lambda$- and $\lambda(t)$-DEOM formalisms are also readily established  for various studies of thermodynamics  (\Sec{thsec4}).
Dissipaton thermofield (DTF) theory is also established along the similar line (\Sec{thsec5}).

As noticed,  
the fermionic dissipaton theories
can also been readily established in a similar manner,  
with $
 \big(\hat f_{k}\hat f_{j}\big)^{\circ}=-\big(\hat f_{j}\hat f_{k}\big)^{\circ}
$ for fermionic dissipaton operators [\cf \Eq{circ_intro}].
Moreover, the fermionic DEOM has been integrated with electronic structure theory for the 
first--principle simulations on 
realistic spintronic systems in experimental measurements.\cite{Ye16608}
These studies include  Kondo transport, magnetic  anisotropy manipulation, spin--polarized scanning
tunneling spectroscopy and so on.\cite{Wan16125114,Wan16154301,Wan182418,Zhu222094}
Furthermore, there would be a
so--called dissipaton embedded quantum master equation formalism, with system--plus--dissipatons being all incorporated into a single master equation.
This provides an alternative formalism for quantum simulations in both bosonic and fermionic scenarios.

It is worth emphasizing that dissipatons are linear bath hybrid  modes.
Nevertheless, the unavoidable backaction of system to environment will result in  simultaneous actions of two or more dissipatons. 
%
The further development of dissipaton theories should take nonlinear environments into account.
For the quadratic bath coupling, we proposed the stochastic--field--dressed DEOM (SFD-DEOM). \cite{Che21174111}
We had also developed the extended Wick's theorem approach to deal with  a model quadratic environment. \cite{Xu17395,Xu18114103,Che22Arxiv2206_14375}
%
%
%
%
Last but not least, it is also anticipated that dissipaton theories discussed in this Perspective would remain essential to  relativistic quantum mechanics of open systems.

\begin{acknowledgments}
Support from
the Ministry of Science and Technology of China (Grant No.\  2021YFA1200103) and the National Natural Science Foundation of China (Grant Nos.\ 22103073 and 22173088)
is gratefully acknowledged.
 \end{acknowledgments}



\end{document}